%% file: langevin.tex
\documentstyle{europhys}
\input euromacr

\input{psfig}
\begin{document}
\euro{}{}{}{}
\Date{}
\shorttitle{G. FRENKEL \etal THE STRUCTURE OF LANGEVIN'S KERNEL ETC.}
%\begin{center}
%{\bf\LARGE The Structure of Langevin's Memory Kernel From
%  Lagrangian Dynamics}\\
%\vspace{.7cm}
%{\bf Gady Frenkel,Moshe Schwartz}\\
%{\em Raymond and Beverly Sackler Faculty of Exact Sciences\\
%School of Physics and Astronomy\\
%Tel Aviv University, Ramat Aviv, 69978, Israel}
%
%\end{center}
\title{The structure of Langevin's memory kernel from Lagrangian dynamics}
\author{G. Frenkel , M. Schwartz}
\institute{ Raymond and Beverly Sackler Faculty of Exact Sciences\\
School of Physics and Astronomy\\
Tel Aviv University, Ramat Aviv, 69978, Israel}
\rec{}{}
\pacs{
\Pacs{66}{10cb}{Diffusion and thermal diffusion}
\Pacs{45}{50-j}{Dynamics and kinematics of a particle and a system of particles}
       }
\maketitle
\begin{abstract}
We obtain the memory kernel of the generalized Langevin equation,
describing a particle interacting with longitudinal phonons in a
liquid. The kernel is obtained analytically at $T=0^oK$ and
numerically at $T>0^oK$. We find that it shows some non-trivial
structural features like negative correlations for some range of time
separations. The system is shown to have three characteristic time
scales, that control the shape of the kernel, and the transition between
quadratic and linear behavior of the mean squared distance (MSD).
Although the derivation of the structure in the memory kernel is
obtained within a specific dynamical model, the phenomenon is shown to
be quite generic.
\end{abstract}

The Generalized Langevin equation is a powerful tool for the study of dynamic
properties of many interesting physical systems.
It is widely believed that the memory kernel is some decaying function
of time, with
no interesting features, except for the characteristic decay time. An
example is the simple Gaussian form used in the literature \cite{wan,rey}.
Some molecular dynamics simulations yield, on the other hand, non
trivial structure of the memory kernel \cite{posch,benj}. A result of
the same nature was obtained analytically by Chow and Hermans
\cite{chow}. Their memory kernel is positive only at t=0 and negative
elsewhere. The derivation assumes, however, that the Brownian particle
interacts with a viscous fluid and therefore it is not a first
principle derivation. Furthermore the final form of the kernel
obviously violates the requirement that its Fourier transform be
positive for all $\omega$.

  The history of attempts to actually derive the memory kernel of a
generalized Langevin equation from a first principle underlying
Lagrangian or Hamiltonian description starts with the paper of
Feynman and Vernon \cite{feynman}. Feynman and Vernon couple the
Brownian particle linearly to a system of harmonic oscillators,
obtaining an exactly solvable problem. Then the memory kernel is
derived from the microscopic parameters. As explained by Schwartz
and Brustein \cite{schwartz}, it is difficult to envisage a situation where the
Brownian particle is not bound to some small region in space, where
a linear coupling could be justified. A number of authors followed in
this direction.
Zwanzig \cite{zwan} and Lindenberg \cite{lind} developed
general formalisms to obtain the memory kernel, but their actual
implementation, in obtaining the memory kernel in terms of the parameters
of the Hamiltonian, is restricted to systems with linear coupling of
the Brownian particle to the degrees of freedom of the thermal bath.
The formalism of Zwanzig, combined with the work of Mori
\cite{mori1,mori2}, has been extensively used in the research that
followed (for examples see: Goodyear and Stratt \cite{good}, Guenza
\cite{guen}, Heppe \cite{hepp} and lee \cite{lee}).
%A more recent work along similar lines is that of Goodyear and Stratt
%\cite{good}.

 It is clear that a microscopic derivation of the
memory kernel is needed, for the cases where the particle is not bound, to determine whether and under what conditions
the kernel has some interesting structure.
 In this paper we derive the memory kernel for a particle interacting
in a realistic way with the longitudinal density waves of the system
in which it is immersed. (The interaction has the same form as the
electron-phonon interaction). We find that the structure of the
memory kernel has interesting features. The most interesting is the
existence of a region of time for which the memory kernel is
negative. Although our result is obtained for a specific model, we
show later that it must be generic.

  Former work by Munakata considers a Lagrangian describing an
impurity interacting with an elastic periodic lattice and obtains
from it the shape of the memory kernel \cite{munakata}. His derivation
is based, however, on the long time form of the MSD, that is linear in
time. To refine his result to be also valid at very short
times, one needs to obtain the full time dependence of the MSD.
In this article we present a method that yields the mean squared
distance and the memory kernel, both
over the full time range. Our starting point is the Lagrangian
describing the interaction of a particle with longitudinal phonons in
a liquid; considered in refs. \cite{schwartz,brus}.

We consider a system of a particle immersed in an infinite idealized liquid and interacting
with it's longitudinal phonons through a general two body interaction,
u(q).

The Lagrangian of the system is \cite{brus}:

\begin{eqnarray}
L =& & \int d\vec{q}\left(\frac{1}{2}
  \frac{m}{\overline{\rho}}\vec{J}(\vec{q})\cdot\vec{J}(-\vec{q})-\frac{1}{2}V( \vec{q})\rho(\vec{q})\rho(-\vec{q})-\mu(\vec{q})\left[\stackrel{\cdot}{\rho}(-\vec{q})-i\vec{q}\cdot\vec{J}(-\vec{q})\right]\right)\\ \nonumber -& &\int d\vec{q}\left(u(\vec{q})\rho(-\vec{q})exp(i\vec{q}\cdot\vec{x})\right)  + \frac{1}{2}M\stackrel{\cdot}{x}^{2}  
\end{eqnarray}
where m is the mass of the liquid particles, $\overline{\rho}$ is its
average number density, $\vec{J}(\vec{q})$ and $\rho(\vec{q})$ are the
Fourier transforms of the current density and the number density
respectively, the
Fourier transform (FT) of the two body (effective) potential between
particles of the liquid is $V(\vec{q})$ and the FT of the interaction
between the Brownian particle and the liquid is $u(\vec{q})$. The
Lagrange multiplier $\mu(\vec{q})$ is introduced to impose the
eq. of continuity, $\vec{x}$ is the coordinate of the
particle and M is its mass.

Assuming that in the absence of the particle the distribution of the
degrees of freedom of the liquid is given by a Gibbs distribution at
temperature T, Brustein {\it et al.} found \cite{brus} that the
particle obeys a generalized
Langevin eq. .% In ref. \cite{brus} a method for numerically obtaining
%the MSD and the kernel was suggested, but never actually applied.
\begin{eqnarray}
M\stackrel{..}{\vec{x}}(t)=-\int_{-\infty}^{t}dt'\gamma(t-t')\stackrel{.}{\vec{x}}(t')+\vec{F}(t)
\end{eqnarray}
The average of the random force $\vec{F}$ is zero, and the force -
force correlation is related to the memory kernel.

\begin{eqnarray}
& &\left\langle\vec{F}\right\rangle=0\\
& &\left\langle\vec{F}(t)\cdot\vec{F}(t')\right\rangle=3K_BT\gamma(t-t')  
\end{eqnarray}
The memory kernel was also given in terms of the interactions,
phonon frequencies, $\omega(q)$ and the MSD.
\begin{eqnarray}
\gamma(t-t')= \frac{1}{3} \int
d \vec{q}\, q^2 \frac{u^2(q)}{V(q)}cos\left[\omega(q)(t-t')\right]e^{-\frac{q^2}{6}\left\langle (\Delta x(t-t'))^2 \right\rangle}
\end{eqnarray}
In ref. \cite{brus} a numerical method for calculating the kernel was suggested.
The numerical method was based on iterations. To calculate the  MSD
one had to evaluate four dimensional integrals, involving Laplace
transforms, a difficult numerical procedure.
In this article, we suggest a slightly different approach that
involves only one dimensional integrations\\

The first step is to obtain the MSD from a generalized Langevin
eq. with a given memory kernel $\gamma$.
To solve eq.(2), we transform it to Fourier space.
We do it by inserting a step function:
\begin{eqnarray}
M\stackrel{..}{\vec{x}}+\int^\infty_{-\infty}\gamma(t-t')\Theta(t-t')\stackrel{.}{\vec{x}}(t')dt'=\vec{F}(t)
\end{eqnarray}
Solving directly for $\stackrel{\wedge}{\vec{x}}(\omega)$, we obtain:
\begin{eqnarray}
\left\langle \Delta x(t)^2
\right\rangle=& &\frac{1}{2\pi}\int^\infty_{-\infty}d\omega\int^\infty_{-\infty}d\omega
'\frac{\langle\stackrel{\wedge}{\vec{F}}(\omega)\cdot\stackrel{\wedge}{\vec{F}}(\omega
  ')\rangle(e^{i\omega t}-1)(e^{i \omega 't}-1)}{(\sqrt{2\pi}i\omega\Gamma
  (\omega)-M\omega^2)(\sqrt{2\pi}i\omega '\Gamma (\omega ')-M\omega
  '^2)},
\end{eqnarray}
where $\Gamma(\omega)$ is the Fourier transform of
$\gamma(t)\Theta(-t)$, and $\stackrel{\wedge}{\vec{F}}(\omega)$ is the
Fourier transform of $\vec{F}(t)$.\\
Using the fluctuation dissipation relation (4):\\$\langle\stackrel{\wedge}{\vec{F}}(\omega)\cdot\stackrel{\wedge}{\vec{F}}(-\omega')\rangle=3K_BT\sqrt{2\pi}\stackrel{\wedge}{\gamma}(\omega)\delta(\omega-\omega')$,
a straightforward calculation yields
\begin{eqnarray}
\left\langle \Delta x(t)^2
\right\rangle=\frac{6K_BT}{\sqrt{2\pi}}\int_{-\infty}^\infty d\omega\frac{\stackrel{\wedge}{\gamma}(\omega)(1-cos(\omega
  t))}{\omega^2((M\omega)^2+2\pi\Gamma(\omega)\Gamma(-\omega))},
\end{eqnarray}
where $\stackrel{\wedge}{\gamma}(\omega)$ is the Fourier transform of $\gamma(t)$.\\

Rescaling the integral on the right hand side of the above by defining
$y=\omega t$, it is clear that the long time dependence of $\left\langle \Delta x(t)^2
\right\rangle$ is linear in t. The short time dependence is quadratic
in t as expected and given by
\begin{eqnarray}
\left\langle \Delta x(t)^2 \right\rangle
=\left[\frac{3K_BT}{\sqrt{2\pi}}\int_{-\infty}^\infty
d\omega\frac{\stackrel{\wedge}{\gamma}(\omega)}{(M\omega)^2+2\pi\Gamma(\omega)\Gamma(-\omega)}\right]t^2\equiv\varphi
t^2
\end{eqnarray}
%\section{The Iterative Method}

The two required quantities $\gamma(t)$ and $\left\langle \Delta x(t)^2
\right\rangle$ are obtained now by an iterative procedure from the two
coupled eqs. (5) and (8). In order to be specific we need,
however, to determine the potentials u(q) and V(q). We take for V(q) a
constant, V(0), corresponding to a $\delta$ function in real space. For
u(q), the potential between the immersed particle and the fluid
particles we take a potential that has a finite range of the order of
the size of the particle. We choose a Gaussian form
$u(0)e^{-a^2q^2}$. The choice of V(q) yields for $\omega(q)$, the
phonon spectrum
\begin{eqnarray}
\omega(q)=\sqrt{\frac{\stackrel{-}{\rho}V(0)}{m}}|q|\equiv c|q|
\end{eqnarray}
where c is the sound velocity.\\
\begin{figure}
\centerline{\psfig{figure=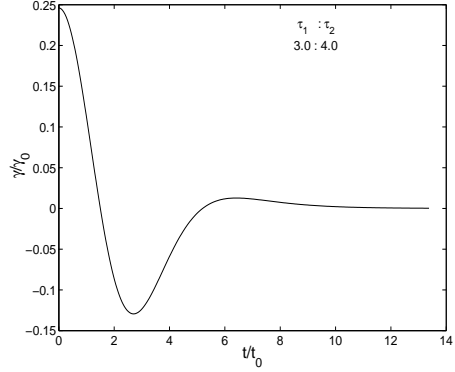,width=6cm,height=5cm,clip=}}

%\vbox to 1cm{\vfill\centerline{\fbox{\psfig{figure=copy_of_fig1-25-4-99.ps,width=10cm,height=10cm,clip=}}\vfill}
%
\caption{ The memory kernel in dimensionless units (
    $\gamma_0\equiv\frac{u(q=0)}{V(q=0)a^5}$.) as a function of
    dimensionless time, for fixed values of $\tau_1$ and $\tau_2$.}
%\label{fig1}
\end{figure}
The iteration procedure is defined by the following eqs.
\begin{eqnarray}
& &G^{(n)}(q,t)= e^{-\frac{1}{6}q^2\langle(\Delta x(t))^2\rangle^{(n)}}\\
& &\gamma^{(n)}(t)= \frac{4\pi}{3}\int dq\,q^4\frac{u^2(q)}{V(q)}cos[\omega(q)(t)]G^{(n)}(q,t)\\
&
&\Gamma^{(n)}(\omega)=\frac{1}{\sqrt{2\pi}}\int_{-\infty}^0\gamma^{(n)}(t)e^{-i
\omega  t}dt\\
& &\stackrel{\wedge}{\gamma}^{(n)}(\omega)=\frac{1}{\sqrt{2\pi}}\int\gamma^{(n)}(t)e^{-i\omega
  t}dt
\end{eqnarray}
and finally
\begin{eqnarray}
\left\langle \Delta x(t)^2
\right\rangle^{(n+1)} = \frac{6K_BT}{\sqrt{2\pi}}\int_{-\infty}^\infty d\omega\frac{\stackrel{\wedge}{\gamma}^{(n)}(\omega)(1-cos(\omega
  t))}{\omega^2((M\omega)^2+2\pi\Gamma^{(n)}(\omega)\Gamma^{(n)}(-\omega))}
\end{eqnarray}
For the specific potentials we chose, the memory kernel is:
\begin{eqnarray}
\gamma(t)=\frac{1}{3}\frac{u^2(0)}{V(0)}\int d\vec{q}q^2
e^{-2a^2q^2}cos\left[ cqt\right]e^{-\frac{1}{6}q^2\left\langle \Delta x(t)^2
\right\rangle}
\end{eqnarray}
Calculating the integral while paying attention to the fact that
$\left\langle \Delta x(t)^2 \right\rangle$ does not depend on q ,we
obtain:
\begin{eqnarray}
\gamma(t)=\frac{\pi u^2(0)}{48V(0)}\frac{\sqrt{\pi}\left(12A^2-12B^2A+B^4\right)e^{-\frac{B^2}{4A}}}{A^{4.5}}
\end{eqnarray}
where:
\begin{eqnarray}
\nonumber  A&=&2a^2+\frac{1}{6}\left\langle \Delta x(t)^2 \right\rangle\\
\nonumber  B&=&ct
\end{eqnarray}
\begin{figure}[ht]
\centerline{\psfig{figure=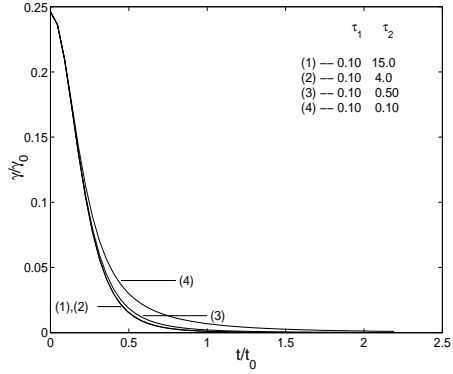,width=6cm ,height=5cm,clip=}}
\caption{Dependence of the memory kernel on $\tau_2$ and time for small $\tau_1$.}
\end{figure}
\begin{figure}
\centerline{\psfig{figure=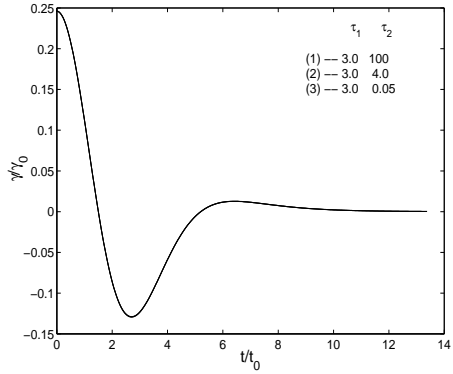,width=6cm
    ,height=5cm,clip=}}
\caption{ Dependence of the memory kernel on $\tau_2$ and time for large
    $\tau_1$. The fig. presents three lines that almost coincide.}
\end{figure}
Since the MSD vanishes with temperature, the calculation yields an
explicit analytic expression for $\gamma(t)$ at $T=0^oK$, with
$A=2a^2$. In ref. \cite{brus} it was found that the friction
coefficient, $\gamma$, vanishes at $T=0^oK$. It is interesting to note
that indeed
\begin{eqnarray}
\int_0^\infty \gamma(t,T=0)dt=0
\end{eqnarray}
The results at finite temperatures can be described in terms of three
natural time scales: $t_0$, $t_1$, and $t_2$. The time $t_0$ is the
time needed by sound to traverse the distance a, that is the size of
the particle; $t_0=\frac{a}{c}$. The time $t_1$ is also related to the
size of the particle and is proportional to the mean time that takes
the particle to move its own size in the short time behavior regime. It is given by $t_1=\frac{a}{\sqrt{\varphi}}$
(eq. 8). The third characteristic time $t_2$ is given by
$t_2=\frac{M}{\gamma}$, where
$\gamma=\stackrel{\wedge}{\gamma}(\omega=0)$ (Notice that the friction
coefficient of the regular Langevin eq. is $\frac{\sqrt{2\pi}}{2}$
  times $\gamma$).
  In fig.(1) we depict a typical kernel that exhibits non-trivial
structure. A fast decay is followed by a negative region, a small
positive peak and then a final decay. The most interesting feature, in
the structure of the memory kernel, is the negative region, of
course. The shape of the memory kernel depends on the two
dimensionless parameters $\tau_1=\frac{t_1}{t_0}$ and
$\tau_2=\frac{t_2}{t_0}$.
 Fig.(2) gives the dependence of the kernel on $\tau_2$ for fixed
$\tau_1<<1$. We see a very weak dependence of the structure on $\tau_2$
ranging from much below to much above unity, without any particular
interesting features. When $\tau_1$ is kept fixed at values above 1
and $\tau_2$ is varied, we see (Fig.(3)) that although the structure
of the memory kernel is very different from the structure for
$\tau_1<<1$ the dependence on $\tau_2$ is still weak.
   In fig.(4) on the other hand, we see the dependence of the kernel on
$\tau_1$ with $\tau_2$ fixed. The striking feature is the transition
from a simple decay at small $\tau_1$ to an interesting structure with
a negative region that becomes more pronounced as $\tau_1$ is
increased. The origin of the negative region can be traced back to the
discussion in refs. \cite{schwartz} and \cite{brus} that shows that at zero temperature 
$\int_0^\infty \gamma(t,T=0)dt$ vanishes. Therefore, it is clear that at zero
temperature $\gamma(t)$ must have a negative region to compensate for
the contribution of the positive region. It is clear therefore, that
for low enough temperature we must also have a negative region. We are
dealing in this paper with a specific model describing a particle
interacting with longitudinal phonons in an idealized liquid. We
would like to stress at this point that the fact that the memory
kernel has a negative region at low temperature is generic and not
just an artifact of the specific model we consider. The reason is that
the fact that at zero temperature $\int_0^\infty \gamma(t,T=0)dt=0$ is
independent of the model and quite general. Therefore, the above
conclusions concerning the existence of a negative region apply to the
general 
case. In fact, this result is supported
by molecular dynamics simulations that show similar behavior of 
the memory kernel \cite{posch,benj}. Returning now to our model, we have to define low enough
temperature in term of a dimensionless quantity.
A careful inspection reveals that the relevant dimensionless parameter
is $\frac{K_BT}{Mc^2}$ and it is easy to show that
$\tau_1=d\cdot\left(\frac{K_BT}{Mc^2}\right)^{-\frac{1}{2}}$.
Indeed d is a function of $\tau_1$ and $\tau_2$, but it does not vanish or diverge as $\tau_1$ or
$\tau_2$ tend to infinity. It is obvious now why when $\tau_1$ is
increased a negative region develops in the kernel. The parameter
$\tau_2$, on the other hand, can be increased without increasing
$\tau_1$ (or equivalently, decreasing $\frac{K_BT}{Mc^2}$). This
explains the weak dependence on $\tau_2$ when $\tau_1$ is fixed.
The physical origin of the negative region is the finite size of the
particle and as result, the interaction of the particle at one point
with phonons emitted earlier at another point. (The emission of
phonons is the mechanism by which the particle looses energy. The
interaction with phonons can contribute to the energy of the particle).
\begin{figure}[ht]
\centerline{\psfig{figure=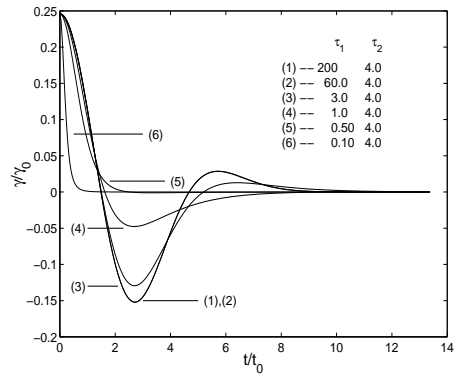,width=6cm ,height=5cm,clip=}}
\caption{ Dependence of the memory kernel on $\tau_1$ and time.}
\end{figure}
\begin{figure}[ht]
\centerline{\psfig{figure=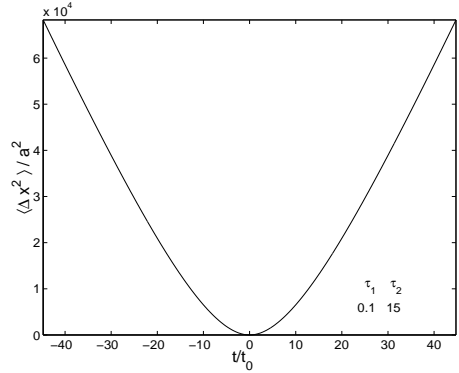,width=6cm ,height=5cm,clip=}}
\caption{Typical MSD, presented in dimensionless units (a is the range of
    interaction between a particle of the medium and the Brownian particle).}
\end{figure}
It is interesting to work out the times t where the kernel
vanishes. This will tell us also when a negative region
exists. We will demand that $\gamma(t)$ given by eq.(17) vanishes and
will assume that $\gamma(t)$ vanishes when t is small enough, so that the
expression to be used for the MSD is $\Delta x^2(t)=\varphi
t^2$ (eq.9). Using the short time form of the MSD is reasonable since at zero temperature
the MSD vanishes, and therefore we are interested at finite temperature
in the region where the MSD is still small.
Solving the system of eqs.
\begin{eqnarray}
\nonumber& &12A^2-12B^2A+B^4=0\\
\nonumber& &A\approx 2a^2+\frac{1}{6}\varphi t^2\\
\nonumber& &B=ct
\end{eqnarray}
we find that the kernel vanishes at two points, one point, or none
at all; depending on $\tau_1$.
\begin{eqnarray}
t(\gamma=0)=\sqrt{\frac{2a^2}{\frac{2}{12\pm\sqrt{96}}c^2-\frac{1}{6}\varphi}}=t_0\sqrt{\frac{2\tau_1^2}{\frac{2}{12\pm\sqrt{96}}\tau_1^2-\frac{1}{6}}}
\end{eqnarray}
When $\tau_1>>1$ a negative region will exist; however, when
$\tau_1<<1$, the negative region will disappear.
In fig.(4), curves (1)-(3) correspond to large $\tau_1$ and indeed the
memory kernel vanishes at two points; the calculation shows that
curves (4) and (5) should have only one zero point as can be seen in
the fig.; and curve (6) should have no negative region at all.

   In fig.(5) we present a typical dependence of the MSD on time. As expected, the MSD is
quadratic for short times, and linear for long times. It
is easy to show, from eq.(8), that the linear long time dependence is $
\left\langle \Delta x(t)^2 \right\rangle
=\frac{12K_BT}{\sqrt{2\pi}\cdot\stackrel{\wedge}{\gamma}(\omega=0)}
\cdot t$, and we have shown the short time dependence to be
$\left\langle \Delta x(t)^2 \right\rangle\equiv\varphi t^2$; therefore
we can easily show that the transition between them occurs at a
typical time scale which depends only on the three known time
scales. When $\tau_1<1$ the transition time is roughly $t_2$.\\

Our calculation is classical and clearly at zero temperature quantum
mechanics prevails. Nevertheless it is obvious that
$\frac{K_BT}{Mc^2}$ can be made small enough while the system still is
not quantum mechanical. Therefore, this generic effect should be experimentally observable.

%\begin{figure}[ht]
%\centerline{\psfig{figure=copy_of_fig1-25-4-99.ps,width=18cm,height=18cm,clip=}}
%\vbox to 1cm{\vfill\centerline{\fbox{\psfig{figure=copy_of_fig1-25-4-99.ps,width=10cm,height=10cm,clip=}}\vfill}
%
%\caption{}
%\label{fig1}
%\end{figure}
%\begin{figure}[ht]
%\centerline{\psfig{figure=copy_of_fig2-25-4-99.ps,width=18cm ,height=18cm,clip=}}
%\caption{}
%\end{figure}
%\begin{figure}
%\centerline{\psfig{figure=copy_of_fig3-26-4-99.ps,width=18cm
%    ,height=18cm,clip=}}
%\caption{}
%\end{figure}
%\begin{figure}[ht]
%\centerline{\psfig{figure=copy_of_fig4-25-4-99.ps,width=18cm ,height=18cm,clip=}}
%\caption{}
%\end{figure}
%\begin{figure}[ht]
%\centerline{\psfig{figure=copy_of_fig5-25-4-99.ps,width=18cm ,height=18cm,clip=}}
%\caption{}
%\end{figure}
%\newpage
%\section{Figure captions}
%
%Fig. (1) :\\
%    The memory kernel in dimensionless units (
%    $\gamma_0\equiv\frac{u(q=0)}{V(q=0)a^5}$.) as a function of
%    dimensionless time, for fixed values of $\tau_1$ and $\tau_2$.\\
%
%Fig. (2) :\\
%    Dependence of the memory kernel on $\tau_2$ and time for small $\tau_1$.
%
%Fig. (3) :\\
%    Dependence of the memory kernel on $\tau_2$ and time for large
%    $\tau_1$. The fig. presents three lines that almost coincide.\\
%
%Fig. (4) :\\
%    Dependence of the memory kernel on $\tau_1$ and time.\\
%
%Fig. (5) :\\
%    Typical MSD, presented in dimensionless units (a is the range of
%    interaction between a particle of the medium and the Brownian particle).
\end{document}

%% file: euromacr.tex
\def\etal{{\hbox{{\tenit\ et al.\/}\tenrm :\ }}}

\newif\ifboo \boofalse